\documentclass[12pt]{article}

\usepackage{amsfonts}
\usepackage{amssymb}
\usepackage{amscd}

\def\D{\hbox{D\kern-.73em\raise.25ex\hbox{-}\raise-.25ex\hbox{ }}}
 \def\d{\hbox{d\kern-.33em\raise.75ex\hbox{-}\raise-.75ex\hbox{}}}

\def\GGG{\frak G }
\def\gr3{\GGG\,(\SSS_3)}

\def\gr2{\GGG\,(\SSS_2)}

\def\SSS{\frak S}

\def\ed{\end{document}}
\def\beq{\begin{equation}}
\def\eeq{\end{equation}}
\def\bea{\begin{eqnarray}}
\def\eea{\end{eqnarray}}
\def\ba{\begin{array}}
\def\ea{\end{array}}
\def\bi{\begin{itemize}}
\def\ei{\end{itemize}}

\newcommand{\bp}{\noindent\begin{minipage}[c]}
\newcommand{\ep}{\end{minipage}}

\begin{document}
\title{\bf  Zeta Nonlocal Scalar Fields }

\author{Branko Dragovich\thanks{\,e-mail
address: dragovich@phy.bg.ac.yu} \\ {}\\
\it{Institute of Physics}\\ \it{Pregrevica 118, P.O. Box 57, 11001
Belgrade, Serbia}}

\date {~}
\maketitle
\begin{abstract}
We consider some  nonlocal and nonpolynomial scalar field models
originated from $p$-adic string theory. Infinite number of spacetime
derivatives is determined by the operator valued  Riemann zeta
function through  d'Alembertian $\Box$ in its argument. Construction
of the corresponding Lagrangians $L$ starts with the exact
Lagrangian $\mathcal{L}_p$ for effective field of $p$-adic tachyon
string, which is  generalized replacing $p$ by arbitrary natural
number $n$ and then taken a sum of $\mathcal{L}_n$ over all $n$. The
corresponding new objects we call zeta scalar  strings. Some basic
classical field properties of these fields are obtained and
presented in this paper. In particular, some solutions of the
equations of motion and their tachyon spectra are studied. Field
theory with Riemann zeta function dynamics is interesting in its own
right as well.
\end{abstract}
\bigskip

{\it \hskip5cm Dedicated to  Vasiliy Sergeevich Vladimirov}

{\it \hskip5cm on the occasion of his 85th birthday}

\section{Introduction}

Already two decades have passed since the first paper on a $p$-adic
string was published \cite{volovich1}. So far various $p$-adic
structures have been observed not only in string theory but also in
many other models of modern mathematical physics (for a review of
the early days developments, see e.g. \cite{freund1,volovich2}).

One of the greatest achievements in $p$-adic string theory is an
effective field description of open scalar $p$-adic strings
\cite{freund2,frampton1}. The effective tachyon Lagrangian is very
simple and exact. It describes not only four-point scattering
amplitudes but also all higher ones at the tree-level.

This $p$-adic string theory  has been significantly  pushed forward
when was shown \cite{sen} that it describes tachyon condensation and
brane descent relations simpler than by ordinary bosonic strings.
After this success, many aspects of $p$-adic string dynamics have
been investigated and compared with dynamics of ordinary strings
(see, e.g. \cite{minahan,sen1,zwiebach,arefeva1} and references
therein). Noncommutative deformation of $p$-adic string world-sheet
with a constant B-field was  investigated in \cite{ghoshal-grange}
(on $p$-adic noncommutativity see also \cite{dragovich1}). A
systematic mathematical study of spatially homogeneous solutions of
the relevant nonlinear differential equations of motion has been of
considerable interest (see \cite{zwiebach,vladimirov1,vladimirov2,
barnaby1} and references therein). Some possible cosmological
implications of $p$-adic string theory have been also investigated
\cite{arefeva2,arefeva3,barnaby2,arefeva4,calcagni}. It was proposed
\cite{ghoshal} that $p$-adic string theories provide lattice
discretization to the world-sheet of ordinary strings. As a result
of these developments, some nontrivial features of ordinary string
theory have been reproduced from the $p$-adic effective action.
Moreover, there have been established so far many similarities and
analogies between $p$-adic and ordinary strings.

Adelic approach to the string scattering amplitudes is a very useful
way to connect $p$-adic and ordinary counterparts (see
\cite{freund1, volovich2} as a review). Moreover, it eliminates
unwanted prime number parameter $p$ contained in $p$-adic amplitudes
and also cures the problem of $p$-adic causality violation. Adelic
generalization of quantum mechanics was also successfully
formulated, and it was found a connection between adelic vacuum
state of the harmonic oscillator and the Riemann zeta function
\cite{dragovich2}. Recently, an interesting approach  toward
foundation of a field theory and cosmology based on the Riemann zeta
function was proposed in \cite{volovich3}. Note that $p$-adic and
ordinary sectors of the four point adelic string amplitudes
separately contain the Riemann zeta function (see, e.g.
\cite{freund1}, \cite{volovich2} and \cite{dragovich4}).

Main motivation for the present paper is our intention to obtain the
corresponding effective Lagrangian for  adelic scalar string. Hence,
as a first step we investigate possibilities to derive Lagrangian
related to the $p$-adic sector of adelic string. Starting with the
exact Lagrangian for the effective field of $p$-adic tachyon string,
extending prime number $p$  to arbitrary natural number $n$ and
undertaking various summations of such Lagrangians over all $n$, we
obtain some scalar field theories with the operator valued Riemann
zeta function. Emergence of the Riemann zeta function at the
classical level can be regarded as its a counterpart in quantum
scattering amplitude. As we shall see this zeta function controls
spacetime nonlocality. In the next sections we construct and explore
some classical field models which may be regarded as candidates for
description of some properties of an adelic open scalar string.

\section{Modeling of some zeta nonlocal scalar fields}

The exact tree-level Lagrangian of effective scalar field $\varphi$,
which describes open $p$-adic string tachyon, is

\beq {\cal L}_p = \frac{m_p^D}{g_p^2}\, \frac{p^2}{p-1} \Big[
-\frac{1}{2}\, \varphi \, p^{-\frac{\Box}{2 m_p^2}} \, \varphi  +
\frac{1}{p+1}\, \varphi^{p+1} \Big]\,,  \label{2.1} \eeq where $p$
 is any prime number, $\Box = - \partial_t^2  + \nabla^2$ is the
$D$-dimensional d'Alembertian and we adopt metric with signature
$(- \, + \, ...\, +)$. An infinite number of spacetime derivatives
follows from the expansion
$$
p^{-\frac{\Box}{2 m_p^2}} = \exp{\Big( - \frac{1}{2 m_p^2} \ln{p}\,
\Box \Big)} = \sum_{k = 0}^{+\infty} \, \Big(-\frac{\ln p}{2 m_p^2}
\Big)^k \, \frac{1}{k !}\, \Box^k \,.
$$
The equation of motion for (\ref{2.1}) is

\beq p^{-\frac{\Box}{2 m_p^2}}\, \varphi = \varphi^p \,, \label{2.2}
\eeq and its properties have been studied by many authors (see e.g.
\cite{zwiebach,vladimirov1,vladimirov2,barnaby1} and references
therein).

It is worth noting that prime number $p$ in (\ref{2.1}) and
(\ref{2.2}) can be replaced by any natural number $n \geq 2$ and
such expressions  also make sense. Moreover,  when $p = 1 +
\varepsilon \to 1$  there is the limit of (\ref{2.1})

\beq {\cal L} = \frac{m^D}{g^2}\,  \Big[ \frac{1}{2}\, \varphi \,
\frac{\Box}{m^2} \, \varphi  + \frac{\varphi^2}{2}\, ( \ln
\varphi^{2} -1 ) \Big]\, \label{2.3}\eeq
 which is related to the ordinary bosonic string in the boundary string field
theory \cite{gerasimov}.

Now we want to introduce a model which incorporates all the above
string Lagrangians (\ref{2.1}) with $p$ replaced by $n \in
\mathbb{N}$. To this end, we take the sum of all Lagrangians ${\cal
L}_n$  in the form

\bea L =   \sum_{n = 1}^{+\infty} C_n\, {\cal L}_n   =  \sum_{n=
1}^{+\infty} C_n \frac{ m_n^D}{g_n^2}\frac{n^2}{n -1} \Big[
-\frac{1}{2}\, \phi \, n^{-\frac{\Box}{2 m_n^2}} \, \phi +
\frac{1}{n + 1} \, \phi^{n+1} \Big]\,, \label{2.4} \eea whose
explicit realization depends on particular choice of coefficients
$C_n$, masses $m_n$ and coupling constants $g_n$. To avoid a
divergence problem of  $1/(n-1)$ when $n = 1$ one has to take that
${C_n\, m_n^D}/{g_n^2}$ is proportional to $n -1$. In this paper we
shall consider a case when coefficients $C_n$ are proportional to
$n-1$, while masses $m_n$ as well as coupling constants $g_n$ do not
depend on $n$, i.e. $ m_n =  m , \,\, g_n = g$. Since this is an
approach towards effective Lagrangian  of an adelic string there is
a sense to take mass and coupling constant independent on particular
$p$ or $n$. Namely, it seems to be natural that an adelic physical
object has fixed rational valued parameters. To emphasize that
Lagrangian (\ref{2.4}) describes a new field, which is different
from a particular $p$-adic one, we introduced notation $\phi$
instead of $\varphi$. The two terms in (\ref{2.4}) with $n = 1$ are
equal up to the sign, but we remain them because they provide the
suitable form of total Lagrangian $L$.

We shall consider a simple case

\beq C_n = \frac{n-1}{n^{2+h}} \,, \label{2.5} \eeq where $h$ is a
real number. The corresponding Lagrangian reads

\bea L_{h} =   \frac{m^D}{g^2} \Big[ - \frac{1}{2}\, \phi \,
\sum_{n= 1}^{+\infty} n^{-\frac{\Box}{2 m^2} -h} \, \phi  + \sum_{n=
1}^{+\infty} \frac{n^{-h}}{n + 1} \, \phi^{n+1} \Big] \label{2.6}
\eea and it depends on parameter $h$.

According to the famous Euler product formula one can write
$$ \sum_{n= 1}^{+\infty} n^{-\frac{\Box}{2\, m^2}  - h} = \prod_p \frac{1}{ 1 -
p^{-\frac{\Box}{2\, m^2}   - h}}\,. $$  Recall that standard
definition of the Riemann zeta function is

\beq  \zeta (s) = \sum_{n= 1}^{+\infty} \frac{1}{n^{s}} = \prod_p
\frac{1}{ 1 - p^{- s}}\,, \quad s = \sigma + i \tau \,, \quad \sigma
>1\,, \label{2.7} \eeq which has analytic continuation to the entire
complex $s$ plane, excluding the point $s=1$, where it has a simple
pole with residue 1. Employing definition (\ref{2.7}) we can rewrite
(\ref{2.6}) in the form

 \beq L_{h} = \frac{m^D}{g^2} \Big[ \, -
\frac{1}{2}\,
 \phi \,  \zeta\Big({\frac{\Box}{2\, m^2}  +
h }\Big) \, \phi  +   \sum_{n= 1}^{+\infty} \frac{n^{ - h}}{n + 1}
\, \phi^{n+1} \Big]\,. \label{2.8} \eeq Here
 $\zeta\Big({\frac{\Box}{2\, m^2} + h}\Big)$ acts as a
pseudodifferential operator  \beq \label{2.9}
\zeta\Big({\frac{\Box}{2\, m^2}  + h }\Big)\, \phi (x) =
\frac{1}{(2\pi)^D}\, \int e^{ ixk}\, \zeta\Big(-\frac{k^2}{2\, m^2}
 + h \Big)\, \tilde{\phi}(k)\,dk \,,
 \eeq
where $ \tilde{\phi}(k) =\int e^{(- i kx)} \,\phi (x)\, dx$ is the
Fourier transform of $\phi (x)$. Lagrangian $L_0 $, with the
restriction on momenta $-k^2 = k_0^2 -\overrightarrow{k}^2
> (2 - 2 h)\, m^2 $ and field $|\phi | < 1$, is analyzed in \cite{dragovich3}. In the
sequel we shall consider Lagrangian (\ref{2.8}) with analytic
continuations of the zeta function and the power series $\sum
\frac{n^{-h}}{n + 1} \, \phi^{n+1}$, i.e.

\beq L_{h} = \frac{m^D}{g^2} \Big[ \,- \frac{1}{2}\,
 \phi \,  \zeta\Big({\frac{\Box}{2 \, m^2}  +
h }\Big) \, \phi    + {\cal{AC}} \sum_{n= 1}^{+\infty} \frac{n^{-
h}}{n + 1} \, \phi^{n+1} \Big]\,, \label{2.10} \eeq where
$\mathcal{AC}$ denotes analytic continuation.

 Nonlocal dynamics of this field $\phi$ is
encoded in the pseudodifferential form of the Riemann zeta function.
When the d'Alembertian is in the argument of the Riemann zeta
function we shall say that we have zeta nonlocality. Consequently,
the above $\phi$ is a zeta nonlocal scalar field.

Potential of the above zeta scalar field (\ref{2.10}) is equal to $-
L_h$ at $\Box = 0$, i.e.

\beq V_{h} (\phi) = \frac{m^D}{ g^2}\,\Big( \frac{\phi^2}{2} \,
\zeta ( h)   - \mathcal{AC} \sum_{n= 1}^{+\infty} \frac{n^{ - h}}{n
+ 1}\, \phi^{n +1} \Big)\,, \label{2.11}\eeq where $h \neq 1$ since
$\zeta (1) = \infty$. The term with $\zeta$-function vanishes at $h
= -2, -4, -6, \cdots$.

The equation of motion in differential and integral form is

\bea  \zeta\Big({\frac{\Box}{2\, m^2}  + h }\Big) \, \phi
= \mathcal{AC} \sum_{n = 1}^{+\infty} n^{- h}\, \phi^n \,, \\
\frac{1}{(2\pi)^D}\, \int_{\mathbb{R}^D} e^{ ixk}\,
\zeta\Big(-\frac{k^2}{2 \, m^2}  + h \Big)\, \tilde{\phi}(k)\,dk =
\mathcal{AC} \sum_{n = 1}^{+\infty} n^{ - h}\, \phi^n \,,
\label{2.12} \eea respectively. Obviously $\phi =0$ is a trivial
solution for any real $h$. Existence of other trivial solutions
depends on parameter $h$. When $h > 1$ we have another trivial
solution $\phi = 1$.

In the weak field approximation $(|\phi (x)|\ll 1)$ the above
expression (\ref{2.12}) becomes

\beq  \int_{\mathbb{R}^D} e^{i k x} \, \Big[\zeta\Big(-\frac{ k^2}{2
\, m^2} + h \Big) - 1 \Big]\, \tilde{\phi}(k)\, dk = 0 \,,
\label{2.13} \eeq which has a solution $\tilde{\phi}(k) \neq 0$ if
equation

\beq \zeta\Big( \frac{-k^2}{2\, m^2} + h \Big) = 1\,, \label{2.14}
\eeq is satisfied. Taking usual relation $k^2 = - k_0^2
+\overrightarrow{k}^2 = - M^2$ equation (\ref{2.14}) in the form

\beq \zeta\Big( \frac{M^2}{2\, m^2} + h \Big) = 1\,, \label{2.15}
\eeq determines mass spectrum $ M^2 = \mu_h \, m^2$, where set of
values of spectral function $\mu_h$ depends on $h$.

Equation (\ref{2.15}) gives infinitely many tachyon  mass solutions.
Namely, function $\zeta (s)$ is continuous for real $s \neq 1$ and
changes sign crossing its zeros  $s = - 2n, \, n \in \mathbb{N}$.
 According to relation $\zeta (1 - 2n) = - B_{2n}/ (2n)$ and values
 of the Bernoulli numbers $(B_0 = 1, \, B_1 = - 1/2, \, B_2 =  1/6, \, B_4 = - 1/30, \,
 B_6 = 1/42, \, B_8 = - 1/30, \, B_{10} =  5/66, \, B_{12} = - 691/2730, \, B_{14} =  7/6, \,
 B_{16} = - 3617/510, \, B_{18} = 43867/798, \, \cdots )$ it follows
 that $|\zeta (1 - 2n)| = | B_{2n}/ (2n)| > 1$ if and only
 if $n \geq 9$. Taking into account also regions where $\zeta (1 -2n)  >
 0$ we conclude that $\zeta (s) =1$ has two solutions  when $  -20 - 4j < s < - 18 - 4j
 $ for every $j = 0, 1, 2, \cdots$. Consequently, for any $h \in \mathbb{Z}$, we obtain
 infinitely many tachyon masses $ M^2$:
 \beq M^2 = - (40 + 8 j + 2 h - a_j)\, m^2 \quad \mbox{and} \quad M^2 = - (36 + 8 j + 2 h + b_j)\, m^2 ,   \eeq
where $a_j \ll 1$, $b_j \ll 1$ and $j = 0, 1, 2, \cdots$.

\section{Discussion with respect to some $ h$ }

Among formally infinitely many possible values of $h$ in (\ref{2.5})
we are going now to  consider five of them ( $h = 0, \, h = \pm 1$
and $ h = \pm 2$ ), which seem to be the most interesting. The case
$h = -2$ is the simplest form of coefficients $C_n$ which contain $n
-1$. In the case $h = -1$, coefficients $C_n = \frac{n-1}{n} \to 1 $
for large $n$ and Lagrangians $\mathcal{L}_n$ are taken into account
almost at an equal footing. In the third case, coefficients $C_n =
\frac{n-1}{n^2}$ are inverses of those within ${\cal L}_n$, and
considerably simplify obtained expressions. Cases $ h =  1 $ and $ h
=  2$ are taken into account to have a more complete insight about
behavior of the Lagrangian $L_{h}$ around $h = 0$.

\subsection{Case $ h = - 2$}

The Lagrangian (\ref{2.10}), the corresponding potential and
equation of motion now read respectively:

\beq L_{-2} = \frac{m^D}{g^2} \Big[ \,- \frac{1}{2}\, \phi \,
\zeta\Big({\frac{\Box}{2\, m^2} - 2 }\Big) \, \phi + \frac{2 \phi^2
- \phi}{(1 - \phi)^2}  - \frac{1}{2} \ln (1 - \phi)^2 \Big] ,
\label{2.18} \eeq

\beq V_{-2} (\phi) = \frac{m^D}{g^2} \Big[ \frac{\phi - 2 \phi^2}{(1
- \phi)^2}   + \frac{1}{2} \ln (1 -\phi)^2 \Big] ,\label{2.19} \eeq

\beq \zeta\Big({\frac{\Box}{2\, m^2} - 2 }\Big) \, \phi =
\frac{1}{(2\pi)^D}\, \int_{\mathbb{R}^D} e^{ ixk}\,
\zeta\Big(-\frac{k^2}{2\, m^2} -2 \Big)\, \tilde{\phi}(k)\,dk =
\frac{\phi (\phi + 1)}{(1 - \phi)^3} \,.\label{2.20} \eeq

Potential $V_{-2} (\phi)$ has one local minimum $V_{-2} (-1) \approx
- 0.057 \, \frac{m^D}{g^2}$ and one local maximum $V_{-2} (0) = 0$.
It is singular at $\phi = 1$ (i.e.  $\lim_{\phi \to 1} V_{-2} (\phi)
= - \infty$) and $\lim_{\phi \to \pm \infty} V_{-2} (\phi) = +
\infty$. Equation of motion (\ref{2.20}) has two trivial solutions:
$\phi (x) = 0$ and $\phi (x) = -1$. Solution $\phi (x) = - 1$ can be
also shown taking $\widetilde{\phi} (k) = - \delta (k) \, (2 \pi)^D$
and $\zeta (-2) = 0$ in (\ref{2.20}).

\subsection{Case $h = - 1$}

Respectively, the corresponding Lagrangian, potential and equation
of motion are:

\beq L_{-1} = \frac{m^D}{g^2} \Big[ - \frac{1}{2} \phi\, \zeta \Big(
\frac{\Box}{2\, m^2} - 1\Big)\,\phi + \frac{\phi}{1 - \phi}  +
\frac{1}{2} \ln (1 -\phi)^2 \Big]\,, \label{2.21} \eeq

\beq V_{-1} (\phi) = \frac{m^D}{g^2} \Big[  \frac{ \zeta (  - 1
)}{2}\,\phi^2 - \frac{\phi}{1 - \phi}  - \frac{1}{2} \ln (1 -\phi)^2
\Big]\,, \label{2.22} \eeq

\beq \zeta\Big({\frac{\Box}{2 m^2} - 1 }\Big) \, \phi =
\frac{1}{(2\pi)^D}\, \int_{\mathbb{R}^D} e^{ ixk}\,
\zeta\Big(-\frac{k^2}{2 m^2} -1 \Big)\, \tilde{\phi}(k)\,dk =
\frac{\phi}{(1 - \phi)^2} \,, \label{2.23} \eeq where $\zeta (  - 1
) = - \frac{1}{12}$.

This potential has the following properties: local maximum
$V_{-1}(0) = 0$, $\, \lim_{\phi \to 1-0} V_{-1}(\phi) = - \infty$,
$\, \lim_{\phi \to 1+0} V_{-1}(\phi) = + \infty$, $\,\lim_{\phi \to
\pm \infty} V_{-1}(\phi) = - \infty$ and there is no stable vacuum.
The equation of motion (\ref{2.23}) has a constant trivial solution
only for $\phi (x) = 0$.

\subsection{Case $ h = 0$}

The related Lagrangian is

\beq L_{0} = - \frac{m^D}{g^2} \Big[ \frac{1}{2} \phi\, \zeta \Big(
\frac{\Box}{2\, m^2}\Big)\,\phi + {\phi}  + \frac{1}{2} \ln (1
-\phi)^2 \Big]\,. \label{2.24} \eeq

The corresponding potential is

\beq V_0 (\phi) = \frac{m^D}{ g^2}\,\Big[\,\frac{ \zeta (0)}{2}\,
\phi^2    + \phi +\frac{1}{2} \ln (1 -\phi)^2 \, \Big]\,,
\label{2.25}\eeq where $\zeta (0) = - \frac{1}{2}$. It has two local
maxima: $ V_0 (0) = 0$ and $V_0 (3) \approx 1.443 \frac{m^D}{g^2}$.
There are no stable points and $\lim_{\phi \to 1} V_0 (\phi) = -
\infty\,, \,\, \lim_{\phi \to \pm \infty} V_0 (\phi) = - \infty \,.
$

The equation of motion is

\beq \zeta\Big( \frac{\Box}{2\, m^2}\Big) \phi =
\frac{1}{(2\pi)^D}\, \int_{\mathbb{R}^D} e^{ ixk}\,
\zeta\Big(-\frac{k^2}{2\, m^2} \Big)\, \tilde{\phi}(k)\,dk =
\frac{\phi}{1 - \phi} \,. \label{2.26} \eeq It has two solutions:
$\phi =0$ and $\phi =3$. The solution $\phi =0$ is evident. The
solution $\phi = 3$ follows from the Taylor expansion of the Riemann
zeta function operator

\beq \zeta\Big( \frac{\Box}{2\, m^2}\Big) = \zeta (0)  + \sum_{n
\geq 1} \frac{\zeta^{(n)} (0)}{n!}\, \Big( \frac{\Box}{2\,
m^2}\Big)^n \,, \label{2.27} \eeq as well as from $\widetilde{\phi}
(k) = (2 \pi)^D \, 3\, \delta (k)$.

\subsection{Case $h = 1$}

Analogously to the previous cases, one has

\beq L_{1} =  \frac{m^D}{g^2} \Big[ - \frac{1}{2} \phi\, \zeta \Big(
\frac{\Box}{2\, m^2} + 1 \Big)\,\phi +  \phi + \frac{1 - \phi}{2} \,
\ln (1 -\phi)^2 \Big]\,, \label{2.28} \eeq

\beq  V_{1} (\phi)  = \frac{m^D}{g^2} \Big[ \frac{\zeta (1)}{2}\,
\phi^2  - \phi  - \frac{1 - \phi}{2} \, \ln (1 -\phi)^2\Big] \,,
\label{2.29} \eeq

\beq \frac{1}{(2\pi)^D}\, \int_{\mathbb{R}^D} e^{ ixk}\,
\zeta\Big(-\frac{k^2}{2\, m^2} + 1 \Big)\, \tilde{\phi}(k)\,dk = -
\frac{1}{2} \, \ln (1 - \phi)^2 \,, \label{2.30} \eeq where $ \zeta
(1) = \infty $ gives $V_{1} (\phi) = \infty$.

\subsection{Case $h = 2$}

\beq L_{2} =  \frac{m^D}{g^2} \Big[ - \frac{1}{2} \phi\, \zeta \Big(
\frac{\Box}{2\, m^2} + 2 \Big)\,\phi  -  \frac{1 - \phi}{2} \ln (1
-\phi)^2  - \phi - \phi \int_0^\phi \frac{\ln (1-w)^2}{2 w}\, dw
\Big]\,. \label{2.31} \eeq

 \beq  V_{2} (\phi) =
\frac{m^D}{g^2}\Big[ \frac{\zeta (2)}{2} \phi^2  + \frac{1 -
\phi}{2} \ln (1 - \phi)^2 + \phi + \phi \int_0^\phi \frac{\ln
(1-w)^2}{2 w}\, dw \Big] \,. \label{2.32}\eeq

\beq  \frac{1}{(2\pi)^D}\, \int_{\mathbb{R}^D} e^{ ixk}\,
\zeta\Big(-\frac{k^2}{2\, m^2} + 2 \Big)\, \tilde{\phi}(k)\,dk = -
\int_0^\phi \frac{\ln (1-w)^2}{2 w}\, dw\,. \label{2.33} \eeq

Since holds equality $- \int_0^1 \frac{\ln (1-w)}{w}\, dw\, =
\sum_{n =1}^\infty \frac{1}{n^2} = \zeta (2)$  one has trivial
solution $\phi = 1$ in (\ref{2.33}).

\section{Concluding remarks}
As a first step towards construction of an effective field theory
for  adelic open scalar string, we have derived a few  Lagrangians
which contain all corresponding $n$-adic Lagrangians ($n \in
\mathbb{N}$). As a result one obtains that an infinite number of
spacetime derivatives and related nonlocality are governed by the
Riemann zeta function. Potentials are nonpolynomial. Tachyon mass
spectra are determined by definite equations. $p$-Adic Lagrangians
can be easily restored from a zeta Lagrangian using just an inverse
procedure for its construction.

This paper contains  some basic classical  properties of the
introduced zeta scalar field. There are still many classical aspects
which should  be investigated. One of them is a systematic study of
the equations of motion and  nontrivial solutions. In the quantum
sector it is  desirable to investigate scattering amplitudes and
make comparison with adelic string.

Using the above procedure there is a sense to consider construction
of a Lagrangian for an open-closed zeta string starting from
$p$-adic ones presented in \cite{freund1}. Note that effective
Lagrangian for open-closed $p$-adic strings was also used to analyze
tachyon condensation \cite{schnabl}.

\section*{\large Acknowledgements}
The work on this article was partially supported by the Ministry of
Science, Serbia, under contract No 144032D. The author thanks I. Ya.
Aref'eva, V. S. Vladimirov and I. V. Volovich for useful
discussions, D. Ghoshal for some comments, and P.G.O. Freund for an
encouragement towards adelic approach. This paper was completed
during author's stay in the Steklov Mathematical Institute, Moscow.

 \end{document}